\documentclass[12pt]{iopart}
\usepackage{graphicx}
\usepackage{dcolumn}
\usepackage{bm}
\usepackage{color} 
\usepackage{array}
\usepackage{float}
\usepackage{supertabular}
\usepackage{longtable}
\begin{document}

\newcommand{\naoki}[1]{\textcolor{blue}{\bf [Naoki: #1]}}
\newcommand{\sid}[1]{\textcolor{red}{\bf [#1]}}

\title{Can Partisan Voting Lead to Truth?}
\author{Naoki Masuda$^{1,2}$ and S. Redner$^3$}
\address{$^1$ Graduate School of Information Science and Technology,
The University of Tokyo,
7-3-1 Hongo, Bunkyo, Tokyo 113-8656, Japan}
\address{$^2$ PRESTO, Japan Science and Technology Agency,
4-1-8 Honcho, Kawaguchi, Saitama 332-0012, Japan}
\address{$^3$ Center for Polymer Studies and Department of Physics, Boston University, Boston, MA USA~ 02215}
\ead{masuda@mist.i.u-tokyo.ac.jp,redner@bu.edu}
\begin{abstract}
  We study an extension of the voter model in which each agent is endowed
  with an innate preference for one of two states that we term as ``truth''
  or ``falsehood''.  Due to interactions with neighbors, an agent that
  innately prefers truth can be persuaded to adopt a false opinion (and thus
  be discordant with its innate preference) or the agent can possess an
  internally concordant ``true'' opinion.  Parallel states exist for agents
  that inherently prefer falsehood.  We determine the conditions under which
  a population of such agents can ultimately reach a consensus for the truth,
  a consensus for falsehood, or reach an impasse where an agent tends to
  adopt the opinion that is in internal concordance with its innate
  preference so that consensus is never achieved.

\end{abstract}


\maketitle

There has been considerable recent public attention in the US on the
following question: ``Is President Obama a Muslim?''~ A Pew Research Center
public opinion poll in the US found that 19\% of respondents answered this
question affirmatively in August 2010~\cite{pew}.  Independent of one's
personal views, there are two incontrovertible facts on this matter: (i) the
question ``Is President Obama a Muslim?''  has a definite answer, and (ii)
there is a lack of consensus on the answer to what seems to be a clear-cut
question.

What is the source of this lack of consensus?  Does conflicting evidence
exist?  Do people interpret the same evidence differently?  What are the
roles of the mass media and fellow peers in influencing the beliefs of
individuals?  In this work, we focus on peer influence by formulating a
voter-like model in which we account for the possibility that each agent's
opinion may be in concordance or discordance with an internal and fixed set
of beliefs.  Our model incorporates two distinct mechanisms for an agent to
change its opinion: one is the desire for an individual to be in concordance
with this inviolate belief set, while the second is the tendency to agree
with neighbors to minimize conflict.  Our goal is to shed light on whether
global truth can ultimately emerge in a socially-interacting population where
the desire for internal concordance may conflict with global truth.  Our
approach is in the spirit of work by physicists in applying minimalist
modeling to real social phenomena (see, e.g., \cite{G91,GF07,CFL90,MGR10}).

The dichotomy between one's private beliefs and publicly-expressed opinions
can take many forms~\cite{K95}.  One such example is H. C. Andersen's fairy
tale ``The Emperor's New Clothes''~\cite{A37}, in which the Emperor's absence
of clothes is finally exposed by the innocent remark of young child.  The
phenomenon of the spread of false norms has recently been elucidated by a
simple computational model~\cite{CWM05}.  In a related vein, Asch~\cite{A51}
found that respondents to an uncomfortable question would conform to a
clearly false consensus judgment rather than risk the stigma of being viewed
as deviant.  Similarly, white Americans overestimated the degree of support
for forced racial segregation during the 1960's and 70's~\cite{O75}.  While
18\% of whites indicated that they favored segregation, 47\% of these
respondents believed that most whites favored segregation.  Another relevant
example is the maddening feature that the general public has strong
divergences of opinion on issues for which there is agreement within the
scientific community~\cite{KBJ10}.

Motivated in part by these examples, we construct a simple generalization of
the voter model --- the ``partisan'' voter model --- to account for the
competing influences of social consensus and personal concordance.  The
classic voter model accounts for the evolution to consensus in a population
of binary and spineless agents that repeatedly evolve by adopting the opinion
state of a randomly-selected neighbor~\cite{L85}.  In our model, we posit
that each individual has an internal ethos --- namely, a fixed beliefs on a
set of fundamental issues, and that each agent is more likely to alter its
opinion to be in concordance with this internal ethos~\cite{F57}.  We also
ascribe a value to the opinion states that we label as ``truth'' ($T$) and
``falsehood'' ($F$).  Individuals evolve by voter model dynamics but the rate
of an update step depends on the direction of the opinion change.  An opinion
change that makes an agent concordant with its internal ethos occurs
preferentially to a change that causes discordance.  As a result of
interactions among neighboring agents by voter model dynamics, an agent that
intrinsically prefers the truth can thus actually have a false opinion.  The
resulting model bears some resemblance to a multi-state voter model of Page
et al.~\cite{PSS07}, in which an agent can self-adjust its vector set of
opinions to be internally consistent in addition to regular voting dynamics.

\begin{figure}[ht]
\begin{center}
\includegraphics[width=0.3\textwidth]{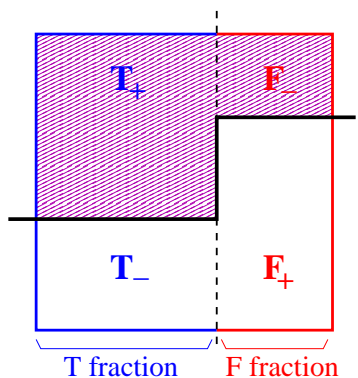}
\caption{ Illustration of the densities of the four types of agents, $T_+$,
  $F_+$, $T_-$, and $F_-$.  The shaded region shows the fraction of the
  population in the true state; this includes a fraction $T_+$ of
  concordant agents that intrinsically prefer the truth and a fraction $F_-$
  of discordant agents that intrinsically prefer falsehood but happen to
  agree with the truth.}
\label{phases}
\end{center}
\end{figure}

In general, there are four possibilities for the opinion state of each agent
(figure~\ref{phases}):
\begin{itemize}
\itemsep -0.35ex
\item An agent intrinsically prefers truth and is in the true state.  Such an
  agent is {\em concordant}.
\item An agent intrinsically prefers falsehood and is in the
  false state. Such an agent is also concordant.
\item An agent intrinsically prefers truth but is in the false state.  Such
  an agent is {\em discordant}.
\item An agent intrinsically
prefers falsehood but is in the true state. Such an agent is also discordant.
\end{itemize}
We define the densities of agents in these four states as $T_+$, $F_+$,
$T_-$, and $F_-$, respectively, each of which evolves with time $t$.  The
subscripts $+$ and $-$ indicate concordance and discordance, respectively.
The total density of agents that intrinsically believe the truth (independent
of their instantaneous state) is the time-independent quantity $T=T_++T_-$
(figure~\ref{phases}), and similarly the density of agents that intrinsically
believe falsehood is $F=F_++F_-$.  However, the density of agents that happen
to be in the true state at a particular time is time-dependent quantity
$T_++F_-$; this includes concordant agents that prefer truth and discordant
agents that prefer falsehood but happen to be in the true state.  Similarly,
the density of agents that happen to be in the false state is $F_++T_-$.  The
total density of agents of any type, {\it i.e.}, $T+F$, is fixed and may be
set to 1 without loss of generality.  Our goal is to understand if global
truth can emerge by this voting dynamics.

We define the evolution of the agents by the following extension of the voter
model: first, an agent is picked at random and then one of its neighboring
agents is picked.  When these two agents are in opposite opinion states
(truth or falsehood), the initial agent is updated with the following rates:
\begin{itemize}
\itemsep -0.5ex
\item A $T$ agent that happens to have a false opinion changes to the true
  (concordant) state with rate $1+\epsilon_t$; a $T$ agent that happens to
  have a true opinion changes to the false (discordant) state with rate
  $1-\epsilon_t$.
\item An $F$ agent that happens to have a true opinion changes to the false
  (concordant) state with rate $1+\epsilon_f$; an $F$ agent that happens to
  have a false opinion changes to the true (discordant) state with rate
  $1-\epsilon_f$.
\end{itemize}
That is, an agent changes to a concordant state with enthusiasm (as
quantified by the enhanced rates $1+\epsilon_t$ or $1+\epsilon_f$ for
changing to the $T$ or $F$ states, respectively, where
$\epsilon_t,\epsilon_f>0$), while an agent changes to a discordant state with
reluctance (reduced rates $1-\epsilon_t$ or $1-\epsilon_f$).

To understand the evolution of the population by this partisan voter dynamics
we first study the mean-field approximation, where all agents can be viewed
as neighbors of each other.  This approximation provides a useful starting
point for investigating the dynamics of the model on more realistic social
networks.  In this mean-field description, the densities of agents in each of
the four possible states evolve by the rate equations:

\begin{equation}
\eqalign{\dot T_+= (1+\epsilon_t)T_-\left[T_++F_-\right] -(1-\epsilon_t)T_+\left[T_-+F_+\right]\,,\cr
\dot T_-= (1-\epsilon_t)T_+\left[T_-+F_+\right] -(1+\epsilon_t)T_-\left[T_++F_-\right]\,,\cr
\dot F_+= (1+\epsilon_f)F_-\left[F_++T_-\right] -(1-\epsilon_f)F_+\left[F_-+T_+\right]\,,\cr
\dot F_-= (1-\epsilon_f)F_+\left[F_-+T_+\right] -(1+\epsilon_f)F_-\left[F_++T_-\right]\,,}
\label{RE}
\end{equation}
where the overdot denotes time derivative.  The terms in \eref{RE} have
simple meanings.  For example, in the equation for $\dot T_+$, the first term
on the right side accounts for the increase (with rate $1+\epsilon_t$) in the
density of concordant agents that are in the true state.  Such a change
arises when a discordant agent that prefers truth interacts with an agent
that currently is in the true state --- either an agent that prefers truth
and is concordant with this internal preference, or an agent that prefers
falsehood but happens to be discordant with its internal preference.  The
equation for $\dot T_-$ can be understood similarly.  The equations for $\dot
F_\pm$ are the same as those for $\dot T_\pm$ by the transformation
$T_\pm\leftrightarrow F_\pm$ and $t\leftrightarrow f$.  Notice, as expected,
that the densities of agents that intrinsically prefer truth or falsehood are
constant, that is, $\dot T=\dot T_++\dot T_-=0$ and similarly $\dot F=\dot
F_++\dot F_-=0$.

Without loss of generality and as one might reasonably expect, we assume that
a majority of the agents intrinsically prefer the truth; that is, $T\geq
\frac{1}{2}$.  To analyze the rate equations \eref{RE}, we introduce the
linear combinations $S=T_++F_+$ (that ranges between 0 and 1), and
$\Delta=T_+-F_+$ (that ranges between $T-1=-F$ and $T$).  Using
$T_+=\frac{1}{2}(S+\Delta)$ and $F_+=\frac{1}{2}(S-\Delta)$, as well as
$T_-=T-T_+$ and $F_-=F-F_+$, we rewrite the rate equations \eref{RE} in terms
of $\Delta$, $S$, and $T$.  Straightforward algebra leads to
\numparts
\begin{eqnarray}
\fl \dot S=(2+\epsilon_t+\epsilon_f)T(1-T)-(\epsilon_t+\epsilon_f)\Delta^2
+\Delta\left[2(1+\epsilon_t+\epsilon_f)T
-(1+\case{1}{2}\epsilon_t+\case{3}{2}\epsilon_f)\right]\nonumber \\
-S+(\epsilon_t-\epsilon_f)\left[S(T-\Delta)-\case{1}{2}S\right]\,,\\
\fl \dot\Delta=(\epsilon_t+\epsilon_f)(T-\Delta)S+\epsilon_f(\Delta-S)+\nonumber \\
+(\epsilon_t-\epsilon_f)\left[T(1-T)+2T\Delta-\Delta^2-\case{1}{2}(S+\Delta)\right]\,.\label{RE2}
\end{eqnarray}
\endnumparts
To understand the dynamical behavior of these equations, first consider the
symmetric (and simpler) case of $\epsilon_t=\epsilon_f\equiv \epsilon$, in
which each type of agent has the same degree of preference toward internal
concordance and the same reluctance for discordance.  For this special case,
the nullclines of equations~\eref{RE}, $\dot S=0$ and $\dot \Delta=0$,
respectively, are:
\begin{equation}
\eqalign{ S=2(1+\epsilon)T(1-T)-2\epsilon\Delta^2
+(1+2\epsilon)(2T-1)\Delta\equiv f(\Delta)\,,\cr
\left(S-\case{1}{2}\right)\left[\Delta-\case{1}{2}(2T-1)\right]=\case{1}{4}(2T-1)\,.}
\label{null}
\end{equation}

\begin{figure}[ht]
\begin{center}
\includegraphics[width=0.35\textwidth]{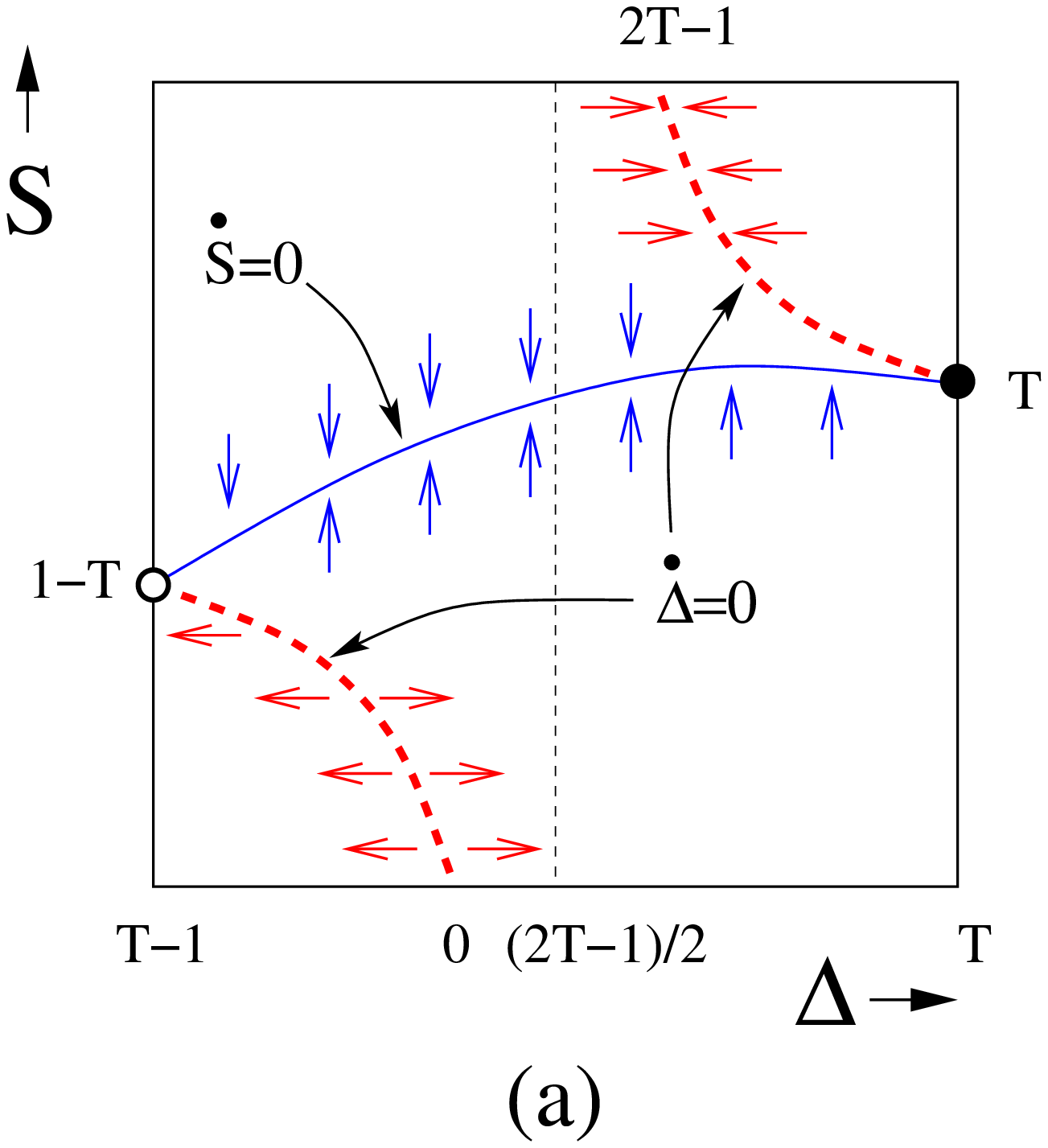}\hskip 1in\includegraphics[width=0.35\textwidth]{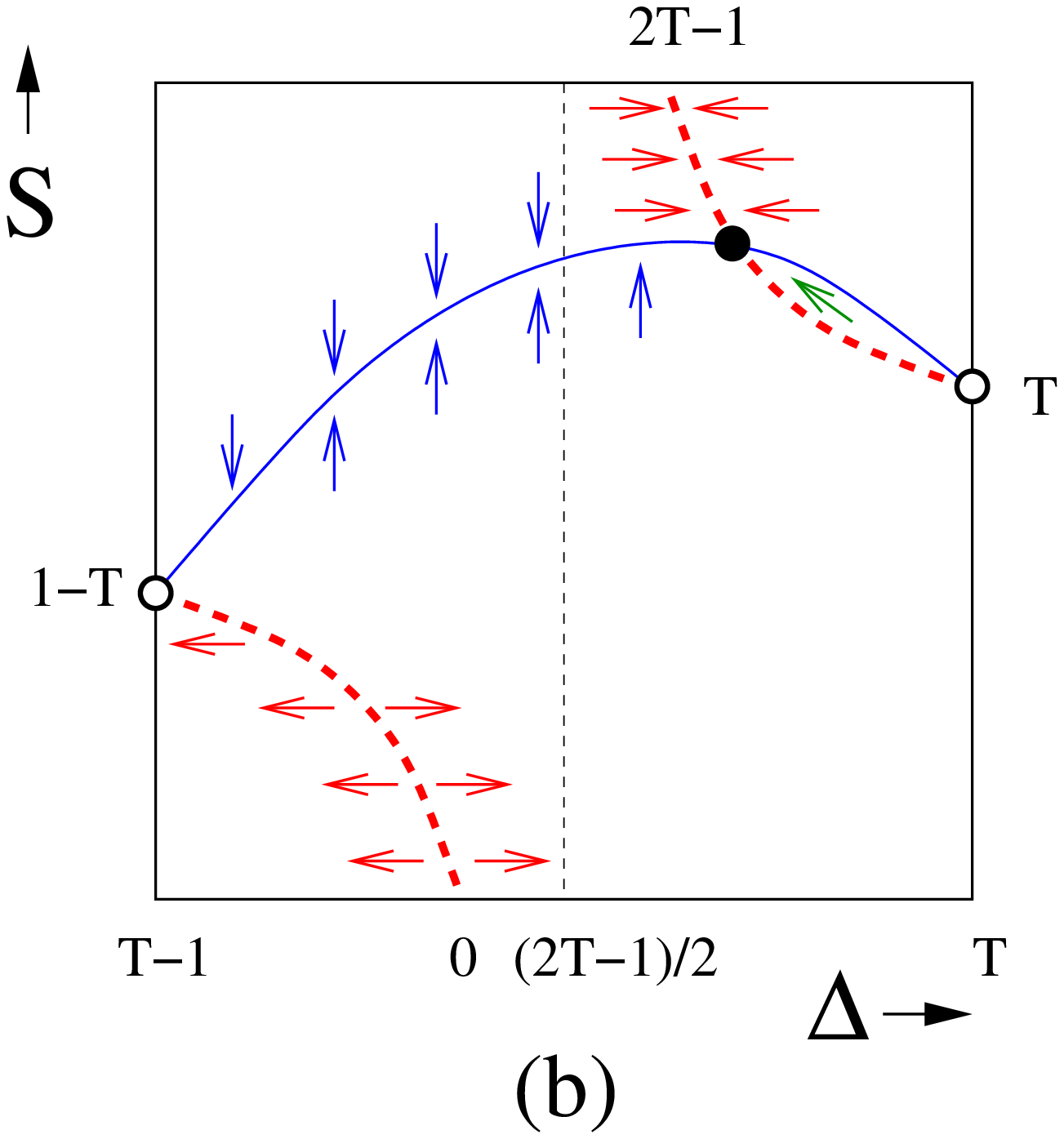}
\caption{Flow diagrams corresponding to the rate equations \eref{RE} when
  $\epsilon_t=\epsilon_f\equiv\epsilon$ in the cases of: (a) weak bias
  $\epsilon<2T-1$ and (b) strong bias $\epsilon>2T-1$.  The nullclines
  defined by $\dot S=0$ (solid) and $\dot\Delta=0$ (dashed) are shown, as
  well as the local flows near each nullcline.  The stable and unstable fixed
  points are denoted by filled and open circles, respectively.  }
\label{flow}
\end{center}
\end{figure}

The locus $S=f(\Delta)$ is monotonically increasing when $\epsilon<2T-1$ but
has a single maximum when $\epsilon>2T-1$, while the locus $\dot\Delta=0$
consists of two hyperbolae in the $\Delta$-$S$ plane (figure~\ref{flow}).
The intersections of these loci give the fixed points of the dynamics.  As
illustrated in figure~\ref{flow}, there are two generic situations.  For weak
intrinsic bias, $\epsilon<2T-1$, there is a stable fixed point at $(\Delta^*,
S^*)=(T,T)$ that corresponds to truth consensus.  In terms of the original
densities $T_\pm$ and $F_\pm$, this fixed point is located at $T_+=T$ and
$F_+=0$.  Thus, truth consensus is reached in which the population consists
of: (i) agents that intrinsically prefer the truth and are concordant, and/or
(ii) agents that intrinsically prefer falsehood but are discordant (see
figure~\ref{phases}).  There is also an unstable fixed point at
$(\Delta^*,S^*)=(T-1,1-T)$ that corresponds to falsehood consensus.  For
$\epsilon>2T-1$, there exists an additional stable fixed point located at
$(\Delta^*,S^*)=\left(\frac{1}{2\epsilon}(1+\epsilon)
  (2T-1),\frac{1}{2}(1+\epsilon)\right)$ in the interior of the $\Delta$-$S$
plane.  This fixed point corresponds to a mixed state of {\em impasse}, where
each agent tends to be aligned with its own internal preference so that
global consensus is not possible.

\begin{figure}[ht]
\begin{center}
\raisebox{0.6cm}{\includegraphics[width=0.275\textwidth]{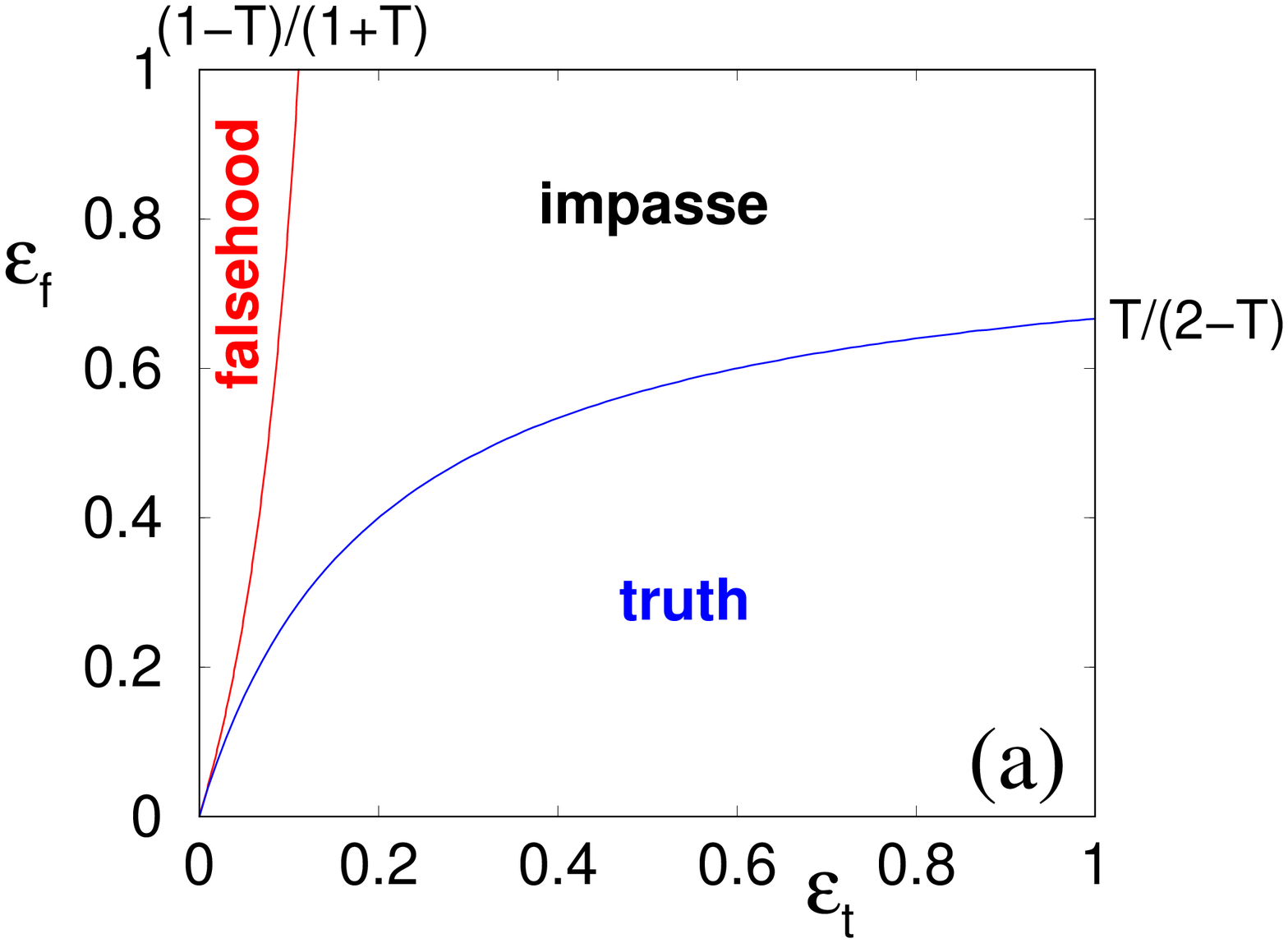}}\quad
\includegraphics[width=0.6\textwidth]{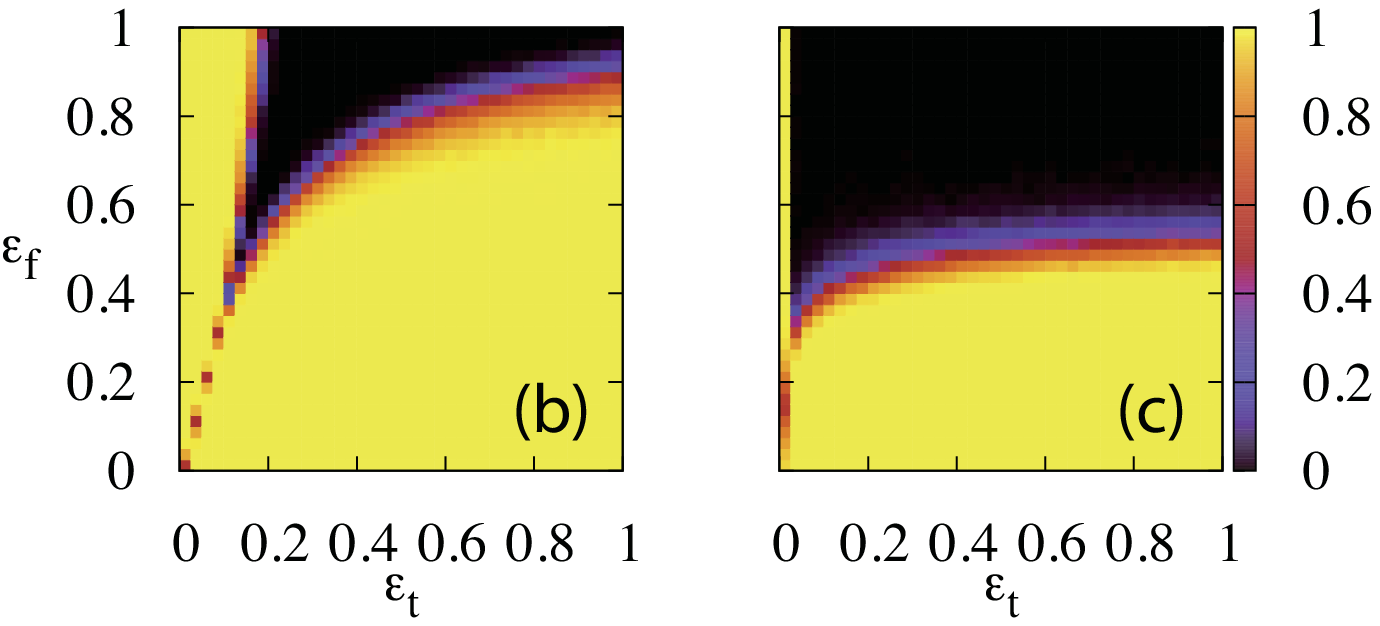}\qquad
\caption{Phase diagrams for the partisan voter model. (a) The attractor state
  of the rate equations (1) as a function of the biases $\epsilon_t$ and
  $\epsilon_f$ when $T=0.8$. (b) Simulation results on a $20\times 20$
  square lattice for the randomly distributed initial state when $T =
  0.8$. (c) Simulation results on a $20 \times 20$ square lattice for a
  clustered initial state of $4\times 4$ agents in the falsehood state.}
\label{bifurcation}
\end{center}
\end{figure}

This same qualitative picture about the nature of the fixed points continues
to hold for general bias parameters $\epsilon_t\ne \epsilon_f$ --- consensus
can be reached only when agents are not too strongly biased toward internal
concordance, while an impasse is reached otherwise.  A striking consequence
of this model is that there is a substantial range for the bias parameters
$\epsilon_t$ and $\epsilon_f$ --- and for values that seem socially realistic
--- for which a consensus of the truth is {\em not\/} reached.  To quantify
this statement, we fix the density of agents that intrinsically prefer the
truth and study the final outcome of the dynamics as a function of
$\epsilon_t$ and $\epsilon_f$.  We make the natural assumption that a
majority of agents intrinsically prefer truth, $T\geq \frac{1}{2}$.  The
transition between impasse and falsehood consensus occurs when the fixed
points corresponding to these two collective states coincide.  This gives the
criterion
\begin{equation}
\Delta^* = T-1 = \frac{T(2\epsilon_t\,\epsilon_f+\epsilon_t+\epsilon_f)
-\epsilon_t\,\epsilon_f-\epsilon_f}{2\epsilon_t\epsilon_f}~.
\end{equation}
Similarly, the transition between impasse and truth consensus is defined by
\begin{equation}
\Delta^* = T = \frac{T(2\epsilon_t\,\epsilon_f+\epsilon_t+\epsilon_f)
-\epsilon_t\,\epsilon_f-\epsilon_f}{2\epsilon_t\epsilon_f}~.
\end{equation}
These two equalities are satisfied when
\numparts
\begin{eqnarray}
\epsilon_f&={T\epsilon_t}/({1-T-\epsilon_t})\,,\\
\epsilon_f&={T\epsilon_t}/({1-T+\epsilon_t})\,,
\end{eqnarray}
\endnumparts
respectively.  These two loci are plotted in figure~\ref{bifurcation}(a) for
the representative case of $T=0.8$.  The region between these two curves
defines the part of the parameter space where the final state is that of
impasse.  Surprisingly (or perhaps not), even when 80\% of the population
intrinsically prefers truth, the impasse state can be reached even when the
bias $\epsilon_t$ toward the truth is stronger than the bias $\epsilon_f$
toward falsehood.  Moreover, this impasse state is reached for a wide range
of bias parameters $\epsilon_t$ and $\epsilon_f$.

To test these predictions and obtain additional empirical insights, we
simulated the partisan voter model on the ring and on the periodic square
lattice.  We studied two natural initial conditions: (i) a random system and
(ii) a droplet state.  In the former, we fix the fraction of voters that
intrinsically believe the truth to be $T=0.8$.  For this population, one-half
of the agents are assigned the true opinion, independent of their intrinsic
beliefs.  In the latter case, the system contains a small cluster of agents
that believe falsehood and also possess the false opinion, while all other
agents believe the truth and possess the true opinion.  We study the
evolution from these two initial states.

Figure~\ref{bifurcation}(b) gives a temperature plot of the fractions of
realizations that reach consensus (either truth or falsehood) within a cutoff
time $5000$ as a function of $\epsilon_t$ and $\epsilon_f$ for the random
initial condition on the square lattice; qualitatively similar results occur
on the ring.  The bright region for small $\epsilon_t$ and large $\epsilon_f$
corresponds to the system reaching falsehood consensus within the cutoff
time.  Conversely, the bright region for large $\epsilon_t$ and small
$\epsilon_f$ corresponds to reaching truth consensus within the cutoff.  In
the dark region, consensus has not yet been reached by the cutoff; we
interpret this behavior as corresponding to impasse.  

The simulation results qualitatively mirror those predicted by the rate
equations in figure~\ref{bifurcation}(a).  However, since consensus must
eventually occur in simulations of a finite system, the lack of consensus by
the cutoff indicates that the consensus time grows exponentially with the
system size in the dark region.  This exponential time dependence is a
generic behavior that will occur whenever the intrinsic voting bias in the
model tends to drive the population away from consensus.  Conversely, in the
portions of the phase diagram that correspond to truth and falsehood
consensus, simulations of a finite system will lead to a consensus time that
generically grows logarithmically with the system size.

Figure~\ref{bifurcation}(c) illustrates the evolution for an initial $4\times
4$ droplet of falsehood (both in belief and in current opinion) in a
background of agents that believe the truth and possess the true opinion;
again, we obtain qualitatively similar results for the corresponding
one-dimensional system.  Although a fraction $T=1-16/400=0.96$ of the agents
prefer the truth and are initially in the true state, impasse (in the sense
outlined above) is the long-time outcome for a substantial range of bias
parameters.  That is, a small fraction of provocateurs --- in the form of
agents that steadfastly believe falsehood --- can forestall the attainment of
consensus.  Moreover, the range of bias parameters for which impasse is
reached is substantial and impasse can arise even when partisans for truth
are more strident than partisans for falsehood.

To summarize, we extended the voter model to include a competition between
conformity --- the tendency to agree with one's neighbors --- and
partisanship --- the desire to be in concordance with one's fixed personal
ethos.  Analysis of the rate equations shows that consensus can be prevented
over a substantial range of model parameters.  A sobering implication of our
model is that a small minority of sufficiently partisan agents that are
inherently predisposed to falsehood can prevent the attainment of consensus
to truth.  It should prove useful to test the validity of the partisan voter
model and its basic parameters from quantitative empirical data.

\ack We thank James Fowler for correspondence that helped motivate this
work, and Damon Centola for discussions and literature advice.  NM
acknowledges financial support by the Grants-in-Aid for Scientific Research
(No.\ 20760258) from MEXT, Japan.  SR acknowledges support from the US
National Science Foundation grant DMR0906504.


\section*{References}
\begin{thebibliography}{99}

\bibitem{pew} http://pewresearch.org/pubs/1701/

\bibitem{G91} S. Galam and S. Moscovici, Eur.\ J. Soc.\ Psych.\ {\bf 21}, 49
  (1991).

\bibitem{GF07} S. Galam and F. Jacobs, Physica A {\bf 381}, 366 (2007).

\bibitem{CFL90} C. Castellano, S. Fortunato, and V. Loreto, Rev.\ Mod.\
  Phys.\ {\bf 81}, 591 (2009).

\bibitem{MGR10} N. Masuda, N. Gibert, S. Redner, Phys.\ Rev.\ E {\bf 82},
  010103(R) (2010).

\bibitem{K95} T. Kuran, {\em Private Truths, Public Lies} (Harvard University
  Press, Cambridge, 1995).

\bibitem{A37} H. C. Andersen, The Emperor’s New Clothes, (Harcourt Brace,
  Orlando, FL, 1837).

\bibitem{CWM05} D. Centola, R. Willer, and M. Macy, ``The Emperor’s Dilemma:
  A Computational Model of Self-Enforcing Norms'', Am.\ J. Sociol.\ {\bf
    110}, 1009--40 (2005).

\bibitem{A51} S. E. Asch, 
  in {\em Groups, Leadership, and Men}, pp.\ 177-–90, ed.\ H. Guetzkow
  (Carnegie Mellon University Press, Pittsburgh, 1951).

\bibitem{O75} H. J. O'Gorman, ``Pluralistic Ignorance and White Estimates of
  White Support for Racial Segregation'', Public Opinion Quarterly {\bf 39}
  313--330 (1975).

\bibitem{KBJ10} D. Kahan, D. Braman, and H. Jenkins-Smith, J. Risk Res.\ {\bf
    14}, 147--174 (2011).

\bibitem{L85} T.~M.~Liggett, {\it Interacting Particle Systems},
  (Springer-Verlag, New York, 1985).

\bibitem{F57} L. Festinger, {\em A Theory of Cognitive Dissonance}, (Row,
  Peterson \& Company, Evanston, IL, 1957).

\bibitem{PSS07} S. Page, L. M. Sander, and C. M. Schneider-Mizell, 
  J. Stat.\ Phys.\ {\bf 128}, 1279 (2007).


\end{thebibliography}
\end{document}